\documentclass[12pt,preprint]{aastex}

\usepackage{epsfig}

\usepackage{xcolor}

\usepackage{soul}

\usepackage{amsmath,amssymb,gensymb}

\begin{document}

\title{Individual Estimates of the Virial Factor in 10 Quasars: Implications on the Kinematics of the Broad Line Region }

\author{E. MEDIAVILLA\altaffilmark{1,2}, J. JIM\'ENEZ-VICENTE\altaffilmark{3,4}, J. MEJ{\'I}A-RESTREPO\altaffilmark{5}, V. MOTTA\altaffilmark{6}, E. FALCO\altaffilmark{7}, J. A. MU\~NOZ\altaffilmark{8,9}, C. FIAN\altaffilmark{1,2}  \& E. GUERRAS\altaffilmark{10}}

\altaffiltext{1}{Instituto de Astrof\'{\i}sica de Canarias, V\'{\i}a L\'actea S/N, La Laguna 38200, Tenerife, Spain}
\altaffiltext{2}{Departamento de Astrof\'{\i}sica, Universidad de la Laguna, La Laguna 38200, Tenerife, Spain}
\altaffiltext{3}{Departamento de F\'{\i}sica Te\'orica y del Cosmos, Universidad de Granada, Campus de Fuentenueva, 18071 Granada, Spain}
\altaffiltext{4}{Instituto Carlos I de F\'{\i}sica Te\'orica y Computacional, Universidad de Granada, 18071 Granada, Spain}
\altaffiltext{5}{European Southern Observatory, Alonso de C\'ordova 3107, Vitacura, Santiago, Chile.}
\altaffiltext{6}{Instituto de F\'{\i}sica y Astronom\'{\i}a, Facultad de Ciencias, Universidad de Valpara\'{\i}so, Avda. Gran Breta\~na 1111, 2360102 Valpara\'{\i}so, Chile}
\altaffiltext{7}{Harvard-Smithsonian Center for Astrophysics, 60 Garden St., Cambridge, MA 02138, USA}
\altaffiltext{8}{Departamento de Astronom\'{\i}a y Astrof\'{\i}sica, Universidad de Valencia, 46100 Burjassot, Valencia, Spain.}
\altaffiltext{9}{Observatorio Astron\'omico, Universidad de Valencia, E-46980 Paterna, Valencia, Spain}        
%\altaffiltext{7}{Departamento de Estad\'{\i}stica e Investigaci\'on Operativa, Universidad de C\'adiz, Avda Ram\'on Puyol s/n, 11202, Algeciras, C\'adiz, Spain}
\altaffiltext{10}{Homer L. Dodge Department of Physics and Astronomy, The University of Oklahoma, Norman, OK, 73019, USA}

\begin{abstract}
Assuming a gravitational origin for the Fe III$\lambda\lambda$2039-2113 redshift and using microlensing based estimates of the size of the region emitting this feature, we  obtain individual measurements of the virial factor, $f$,  in 10 quasars. The average values for the Balmer lines, $\langle f_{H\beta}\rangle={\bf 0.43\pm 0.20}$ and $\langle f_{H\alpha}\rangle={\bf 0.50\pm 0.24}$,  are in good agreement with the results of previous studies for objects with lines of comparable widths.{ In the case of Mg II, consistent results, $f_{Mg II} \sim {\bf 0.44}$, can be also obtained accepting a reasonable scaling for the size of the emitting region}.  The modeling of the cumulative histograms of individual measurements, $CDF(f)$, indicates a {relatively} high value for the ratio between isotropic and cylindrical motions, $a\sim {\bf 0.4}-0.7$. On the contrary, we find very large values of the virial factor associated to the Fe III$\lambda\lambda$2039-2113 blend, $f_{FeIII}=14.3\pm2.4$, which can be explained  if this feature arises from a flattened nearly face-on structure, similar to the accretion disk. 

\end{abstract}


\section{Introduction \label{intro}}

The measurement of the mass of central super massive black holes (SMBH) in quasars is mainly based on the application of the virial theorem to the emitters that give rise to the observed broad emission lines (BEL). The basic idea is to use the Doppler broadening of these lines, $\Delta V$, related to the motion of the emitters under the gravitational pull of the SMBH, to estimate its mass using the virial equation,

\begin{equation}
M_{BH}= f {(\Delta V)^2R\over G}.
\end{equation}
A measurement of the radius of the emitting region, $R$, obtained from Reverberation Mapping (RM) or other alternative method is also needed (see a summary of techniques in Campitiello et al. 2019). This equation is affected by a scaling factor, $f$, which encompasses all the information about the unknown geometry and dynamics (which could include non gravitational forces) of the BLR. Usually we have only statistical information about this factor which, in principle, may be very different from object to object and for different emission lines. For these reasons, $f$ is commonly considered the major source of uncertainty in virial mass determinations. However, the large experimental uncertainties in the estimate of SMBH masses from primary methods, the complexity of the broad emission lines, and the lack of agreement in the definition and systematics of the procedures to estimate the line widths, make this a non obvious conclusion. {In any case, the knowledge of the range of values spanned by $f$ is very important, not only to analize the limitations and possibilities of individual and average mass determinations using the virial, but also to understand the physics of the BLR.}

To estimate the virial factor, $f$, an alternative way\footnote{i.e. independent from the virial.} to measure the mass of the SMBH is needed. This has been done in a number of previous works assuming that AGNs follow the $M_{BH}-\sigma_*$ relation of inactive galaxies (Ferrarese \& Merrit 2000; Gebhardt et al. 2000). The resulting measurements span a large range of values (see the compilation by Campitiello et al. 2019) from $f_{\sigma}=2.8\pm 0.6$\footnote{The value of $f$ is different if we take $\sigma$ or $FWHM$ as the line width indicator. For Gaussian line profiles, $FWHM=2.35 \sigma$, but in many applications of the virial using emission line
profiles,  $\sigma$ is the second moment of the experimental line profile, and the $FWHM/\sigma$ ratio depends on the profile
shape (Collin et al. 2006).} (Onken et al. 2004) to $f_{\sigma}=5.5\pm 1.8$ (Graham et al. 2011), with a large scatter. Part of this scatter may be intrinsic. In fact, some studies (Collin et al. 2006; Ho \& Kim 2014) indicate that AGNs can be separated in populations with different values of $f$. Using mass estimates based on accretion disk fitting, Campitiello et al. (2019) derive $f_\sigma$ for a large sample of objects, obtaining values ($\langle \log f_\sigma\rangle=0.63\pm0.49$), consistent to within uncertainties with other values\footnote{This value can change with the BH spin range considered in the models.} in the literature. However, the large scatter ($f$ spans a range of $\sim2$ orders of magnitude), likely due to the uncertainties in $f$ measurements, makes difficult the interpretation of the results.

A new method to infer SMBH masses (Mediavilla et al. 2018, 2019) based on the measurement of the redshift of the Fe III$\lambda\lambda$2039-2113 blend is now available. This method is free from geometrical effects and largely
insensitive to nongravitational forces. We propose to use it, in combination with microlensing based estimates of the size of the region emitting the Fe III$\lambda\lambda$2039-2113 blend, to determine $f$. We will apply the method to the sample of 10 quasars of Capellupo et al. (2015, 2016) for which redshifts of the Fe III$\lambda\lambda$2039-2113 blend have been measured (Mediavilla et al. 2019), and to the composite quasar spectra of the Baryon Oscillation Spectroscopic Survey (BOSS, Jensen et al. 2016).

The paper is organized as follows. {In \S 2.1. we derive virial factors for the sample of 10 quasars and composite BOSS spectra.  In \S 2.2. we analyze} the experimental correlation between the redshift of the Fe III$\lambda\lambda$2039-2113 blend and the squared widths of several emission lines.  {Section 3. is devoted to} discuss the geometry and the kinematics of the BLR. Finally, the main conclusions are summarized in \S 4.

\section{Results\label{proxy}}

\subsection{Virial Factor Determinations }
\subsubsection{Individual Quasar Spectra from Mej{\'{\i}}a-Restrepo et al. (2016)\label{secind}}

Using the virial theorem, 

\begin{equation}
\label{virial}
M_{BH}\simeq f_{H\beta}{FWHM_{H\beta}^2R_{H\beta}\over G}
\end{equation}
and the equation relating (under the gravitational redshift hypothesis) the SMBH mass with the  redshift of the Fe III UV lines (see Eq. 3 in Mediavilla et al. 2018):

\begin{equation}
\label{mass}
M_{BH}\simeq{2 c^2\over3 G}{\left(\Delta \lambda\over \lambda\right)_{FeIII}}{R_{FeIII}},
\end{equation}
we can obtain the virial factor in terms of the squared widths, the redshifts, and the ratio between sizes,

\begin{equation}
\label{virialfactor}
 f_{H\beta} \simeq{2 \over 3}{R_{FeIII}\over R_{H\beta}}{\left({\Delta \lambda\over \lambda}\right)_{FeIII}\over \left(FWHM_{H\beta}/c\right)^2},
\end{equation}
(notice that here and hereafter we calculate $f$ taking the FWHM as the line width indicator). To estimate $R_{H\beta}$ we can use the size vs. luminosity, R-L,  scaling adopted by Mej{\'{\i}}a-Restrepo et al. (2016)\footnote{The slope of 0.65 adopted for H$\beta$ is motivated by the trend of the slope at the high-luminosity end of Bentz et al. (2013) data, which is suitable for our sample of objects taken from Mejía-Restrepo et al. (2016).}, 

\begin{equation}
\label{RHb}
R_ {H\beta}=538 \left({\lambda L_{\lambda5100}\over 10^{46} {\, \rm erg\,s^{-1}}}\right)^{0.65}\rm light\, days.
\end{equation}
In the case of $R_ {FeIII}$ we can use the average microlensing size estimated by Fian et al. (2018) rescaling it by applying the $R\propto \sqrt{\lambda {L}_{\lambda }}$ relationship from photoionization theory \footnote{Instead of using this relationship directly inferred from microlensing and photoinization theory, we could have used the R-L relationships by Mediavilla et al. (2018), which is quite similar, or by Mediavilla et al. (2019), indirectly derived to match the redshift-based and virial masses obtained using emission lines arising from much larger
regions.}. Inserting the value of $\langle R\rangle $ from Fian et al. (2018) and the average of the square root of the luminosities of the quasars, $\langle \sqrt{\lambda {L}_{\lambda }}\rangle $, used by these authors to infer $\langle R\rangle $, we obtain,

\begin{equation}
\label{RFeIII}
R_ {FeIII}=13.3^{+6}_{-5}  \left({\lambda L_{\lambda1350}\over 10^{45.79} {\,\rm erg\,s^{-1}}}\right)^{0.5}\rm light\, days.
\end{equation}
Using the luminosities from Mej{\'{\i}}a-Restrepo et al. (2016) and Eqs. \ref{virialfactor}, \ref{RHb}, and  \ref{RFeIII}, we compute the virial factors,  $f_{H\beta}$ {(see Table 1)}, for each of the 10 quasars from Mej{\'{\i}}a-Restrepo et al. (2016)  considered in Mediavilla et al. (2019). {In Table 1 we also include the} virial factors corresponding to H$\alpha$ obtained using the same R-L relationship as for H$\beta$. The cumulative {histograms of values are represented} in Figure \ref{Histograms}. The mean values of the distributions are ($1\,\sigma$ uncertainties), $\langle f_{H\beta}\rangle = {\bf 0.43\pm 0.20}$ and $\langle f_{H\alpha}\rangle = {\bf 0.50\pm 0.24}$. 

{To analyze the impact of the choice of the R-L slope in Eqs. \ref{RHb} and  \ref{RFeIII}, we repeat the calculations applying the  widely used relationship, with slope 0.53, found for the entire luminosity range by Bentz et al. (2013), to $R_ {H\beta}$ and the same slope to $R_ {FeIII}$. The results imply a $\sim$ 30\% systematic shift of the virial factors towards higher values. In general, a shift of the  $R_ {H\beta}$ ($R_ {FeIII}$) slope towards smaller (higher) values would result in a systematic increase of the virial factor estimates.}

{We lack on a reliable R-L scaling relationship for Mg II. However, recent results from the SDSS RM project (Shen et al. 2019) indicate that the Mg II emitting region is $\sim 2$ and $\sim$1.4 times greater than the  regions corresponding to C IV and CIII], respectively. The larger size of the region emitting the CIII] line seems reasonable as this line shows less variability and is less prone to microlensing than C IV, which may exhibit a variability comparable to that of the Balmer lines (i.e., $R_{CIV}\sim R_{H\beta}$). Assuming, $R_{MgII}\sim 2R_{CIV}\sim 2R_{H\beta}$, we obtain {the virial factors for MgII, $f_{MgII}$, (see Table 1).  The }mean value for the virial factor $\langle f_{MgII}\rangle \sim {\bf 0.44\pm 0.21}$, {is} consistent with the ones obtained for the Balmer lines. }

\subsubsection{Composite Quasar Spectra from BOSS }
{ {From the values of $FWHM_{MgII}$ and $\left({\Delta \lambda/ \lambda}\right)_{FeIII}$ of BOSS composite spectra (Mediavilla et al. 2018) and Eq. \ref{virialfactor},  we estimate\footnote{For the BOSS composite spectra that include the MgII emission line within the wavelength coverage.} $f_{MgII}$ (see Table 2). We use the same assumption about the size of the MgII region, $R_{MgII}\sim 2R_{CIV}\sim 2R_{H\beta}$. The} average virial factor inferred from  BOSS composites ($\langle f_{MgII}\rangle={\bf 0.35\pm 0.15}$) is consistent with the ones inferred in the previous section from the individual spectra.}

In Mediavilla et al. (2108), we have also derived values of $FWHM_{FeIII}$ for the BOSS quasar composite spectra. Inserting these values in Eq. \ref{virialfactor} (in this case the ratio between sizes is trivially 1) we obtain very high {values} for the virial factor, $ f_{FeIII}$ {(see Table 2),  with mean}, $\langle f_{FeIII}\rangle = 14.3\pm 2.4$. 
%{\color{blue} We can derive a similar estimate of $f_{FeIII}$ from the fit performed in Mediavilla et al. (2018) to $\log\left(\sigma^2/\lambda^2\right)$ vs. $\log \left(\Delta \lambda/ \lambda\right)$. Comparing Eqs. 6 and 7 of Mediavilla et al. (2018) we obtain, $\log \left(3f_\sigma/2\right)=-2.09\pm0.64$, and from this, $f_{FeIII}=15^{+50}_{-12}$. Notice that the errors are much greater now because the linear dependence between $\Delta \lambda$ and  $\sigma^2$ (i.e. the virialization) is not {\it a priori} assumed (i.e., it is a result of the fit).}

\subsection{Correlation between the Fe III$\lambda\lambda$2039-2113 Redshifts and  the  Squared Widths of the Broad Emission Lines}

{If, as it seems to be our case, the virial factor does not change very much from object to object, and there is a proportionality between H$\beta$ and  Fe III$\lambda\lambda$2039-2113 sizes, ${R_{FeIII}\propto  R_{H\beta}}$, Eq. \ref{virialfactor} can be rewritten to express a correlation\footnote{We write the correlation for the case of H$\beta$. Analogous relationships can be written for H$\alpha$ and MgII.} between $FWHM_{H\beta}^2$ and $\left({\Delta \lambda\over \lambda}\right)_{FeIII}$,
\begin{equation}
\label{linear}
\left(FWHM_{H\beta}\over c\right)^2 \simeq{2 \over 3}{1\over f_{H\beta}}{R_{FeIII}\over R_{H\beta}}\left({\Delta \lambda\over \lambda}\right)_{FeIII},
\end{equation}
This correlation has been experimentally found  for  the H$\beta$, H$\alpha$ and Mg II emission lines (Mediavilla et al. 2019) and  for the Fe III$\lambda\lambda$2039-2113  blend (Mediavilla et al. 2018). We can directly fit these correlations to derive statistical estimates of the virial factors for the different lines and compare them with the individual determinations.} {\bf We take the average value for the ratio between sizes, $\langle R_{FeIII}/ R_{H\beta}\rangle$, obtained using Equations \ref{RHb} and \ref{RFeIII}. We also take, $ R_{H\alpha}= R_{H\beta}$ and $ R_{MgII}= 2R_{H\beta}$.}

\subsubsection{Individual Quasar Spectra from Mej{\'{\i}}a-Restrepo et al. (2016)}

Using the sample of quasars of Capellupo et al. (2015, 2016), Mediavilla et al. (2019) found a correlation between the redshift of the Fe III$\lambda\lambda$2039-2113 blend and the squared widths of H$\beta$, H$\alpha$ and Mg II { (see Figure 1 of Mediavilla et al. 2019).  A linear fit to Eq. \ref{linear}  results in values very similar to the ones inferred from the average of the individual estimates: $f_{H\beta}=0.43^{+0.06}_{-0.05}$, $f_{H\alpha}=0.46^{+0.07}_{-0.05}$, $f_{MgII}=0.45^{+0.06}_{-0.05}$.
}

{\subsubsection{Composite Quasar Spectra from BOSS \label{proxy3}}

In Mediavilla et al. (2019) {it is} found (see their Figure 2) that the redshifts vs. squared widths correlation derived from the sample of individual quasar spectra is also very well matched by the  Fe III$\lambda\lambda$2039-2113 redshifts and $FWHM_{MgII}^2$ obtained from the high S/N composite quasar spectra of BOSS.
%{ \color{blue} In Mediavilla et al. (2019) the $FWHM_{MgII}$ was estimated in a direct way, fitting a straight line to the continuum and measuring the width at 50\% of the peak intensity with respect to this continuum. However, it is commonly supposed that the wings of MgII are strongly contributed by a pseudo-continuum arising from different species, particularly iron. Thus, to refine this calculation,  we fit and remove these wings using two Gaussians, one to match the broad wings and other to account for the comparatively narrow MgII emission\footnote{{\color{red} More complex approaches to reproduce and remove the Mg II broad wings are out of the scope of this paper.}}. The correlation of the Fe III$\lambda\lambda$2039-2113 redshifts with the FWHM of the narrower Gaussian that we identify with the MgII emission is even better (see Fig. \ref{Correlations}) than in  Mediavilla et al. (2019). The data points (with a 14\% of global scaling) match perfectly with the results (and linear fit) obtained from Mej{\'{\i}}a-Restrepo et al. (2016) data.} 
In principle, for composite spectra, the good correlation between redshifts and widths do not necessarily imply an invariance of the virial factor, as it may be averaged during the staking process to obtain the composites. However, the good quantitative matching of the relationships corresponding to the individual quasar spectra from Mej{\'{\i}}a-Restrepo et al. (2016) and to the BOSS composites would be virtually impossible if the virial factors were very different among quasars. { A linear fit to Eq. \ref{linear}  results in $f^{BOSS}_{MgII}=0.37^{+0.02}_{-0.01}$.}

Finally, a correlation was also found between the redshifts and squared widths of the Fe III$\lambda\lambda$2039-2113 blend (Mediavilla et al. 2018). This experimental correlation reproduces the expected linear relationship between squared widths and redshifts but does not match the {$\langle FWHM^2/(\Delta \lambda/\lambda)\rangle$ ratio} corresponding to the relationship based on the $FWHM^2_{MgII}$. {In this case the linear fit to Eq. \ref{linear} gives $f^{BOSS}_{FeIII}=13.89^{+0.60}_{-0.56}$.} This is, likely, an evidence of a different origin for the MgII and the Fe III emission.

\section{Discussion}

The mean values obtained for the Balmer lines, $\langle f_{H\beta}\rangle = {\bf 0.43\pm 0.20}$ and $\langle f_{H\alpha}\rangle = {\bf 0.50\pm 0.24}$, {and the estimate for Mg II, $\langle f_{MgII}\rangle \sim {\bf 0.44\pm 0.21}$,} match well the $f\sim 0.52$ value found by Collin et al. (2006) for  objects (Population B according to Sulentic et al. 2000) with $FWHM\ge 4000\,\rm km\, s^{-1}$  when the mean spectrum is used to measure the FWHM\footnote{The average widths of the Balmer lines for the quasars in our sample are: $\langle FWHM_{H\beta}\rangle=4805.7\,\rm km\, s^{-1}$ and $\langle FWHM_{H\alpha}\rangle=4514.0\,\rm km\, s^{-1}$.}. These values are also in agreement with the results by Ho \& Kim (2014) for pseudobulges ($0.5\pm0.2$). Our measurements are consistent with the results of Campitiello et al. (2019), $f=0.77^{+1.61}_{-0.52}$, although the large scatter lessen the interest of the comparison. In any case, in the compilation by Campitiello et al. (2019), our results are in good agreement with the estimates obtained from Grier et al. (2019) data.

\subsection{Restrictions on the Range of Virial Factor Estimates}

{As far as the individual error estimates (Table 1) are comparable with the standard deviations of the samples, we cannot be sure that the range spanned by the virial factors is related to intrinsic variations. On the other hand, assuming normality for the parent distribution, a t-test reject the null hypothesis that the mean of the $f_{H\beta}$ population is greater (smaller) than  {\bf 0.63 (0.23)} with a p-value equal to 0.006 ({\bf 0.006}), i.e., the statistical restriction of values imposed by the data is rather tight. Similar results are derived for $f_{H\alpha}$, and $f_{MgII}$.}

{The range of values spanned by the $f_{H\beta}$, $f_{H\alpha}$, and $f_{MgII}$, virial factors is notably small if we compare it with the scatter of the experimental results from Campitiello et al. (2019), $f=0.77^{+1.61}_{-0.52}$. However, any comparison between both results should be taken with care: while Campitiello et al. (2019) gather virial estimates from several sources, which use a rather heterogeneous  sample of objects and methods, we apply a single method to a quasar sample of limited size, covering a relatively small range in luminosity. Thus, the comparatively narrow range of intrinsic variability that can be inferred from our data, can not be extrapolated, in general.}

Regarding the BLR structure {of our sample of quasars, the relative} invariance of  $f_{H\beta}$,  $f_{H\alpha}$, and $f_{Mg II}$ indicates that the geometry and the dynamics of the distribution of emitters is someway regular among {them}. Thinking in two extreme cases, the small range spanned by the virial factors can be achieved if the {kinematics of the} emitters is more or less isotropic (hence insensitive to inclination) or, alternatively, if the emitters lie in a flattened structure whose orientation is limited to a narrow range of values (possibly due to the observational bias limiting quasars identification to objects with low axial inclination). On the other hand, the relative invariance of $f$ also implies that the impact of non gravitational forces is small or restricted to a {relatively} narrow range of values.

% {\color{red} The virial factors corresponding to MgII are slightly larger, $\langle f_{MgII}\rangle =( 0.9\pm 0.4)(R_{H\beta}/R_{MgII})$, but likely need to be scaled down to take into account that MgII arises from a larger region than H$\beta$.}
\subsection{Isotropic versus Cylindrical Kinematics in the BLR}

{According to the previous discussion,} it is interesting to compare {our results} with the expectations corresponding to isotropic and cylindrical kinematics. In the case of an idealized isotropic configuration of orbits (a shell of circular orbits with all the possible orientations, for instance)  the resulting emission line profile (see Appendix \ref{appb}) is a rectangular function of $FWHM=2\sqrt{GM_{BH}\over R}=2V_{Kep}$. Thus, according to Eq. \ref{virial}, the associated virial factor would be $f_{iso}=1/4$. In the cylindrical case, the Doppler broadening is conservatively bounded by the maximum projected Keplerian velocity so that $FWHM < 2V_{Kep}\sin i$, where $i$ is the inclination ($0^{\rm o}$ is face-on). Consequently, $f_{cyl} > 1/(4\sin^2 i)$. According to the unified scheme (Antonucci \& Miller, 1985), quasars cannot be seen edge on and, in fact, it seems that their inclinations are confined to a narrow range of values not far from 0. If we take, conservatively, $i\le i_0=30^º$, we obtain $f_{cyl}> 1$. 

None of the individual virial factors obtained from the Balmer lines is greater than 1 and, in the average, they are relatively close to $f_{iso}$. Thus, our results {support that the kinematics of the Balmer line emitters has a significant contribution from isotropic motion}. {In the case of MgII we obtain similar results, provided that $R_{MgII}/R_{H\beta} \sim 2$ (see \S \ref{secind}).}

%
%, according to Shen et al. (2019), and that $R_{CIV}\sim R_{H\beta}$.}
%

% {\color{red} The situation is less clear for Mg II with  $\langle f_{MgII}\rangle =(0.9\pm 0.4)(R_{H\beta}/R_{MgII})$ and 2 (of 10 objects) with $f\sim 1.5(R_{H\beta}/R_{MgII})$. But these values should be scaled by the factor $R_{H\beta}/R_{MgII}$, which is likely less tan 1.}

We can also use the simple parameterization proposed by Collin et al. (2006) (Eq. 11 of these authors, see also Decarli et al. 2008) to discuss the degree of isotropy of the emitters {kinematics}. According to this model the line width is given by,

\begin{equation}
\label{collin}
FWHM=(a^2+\sin^2 i)^{1/2}2V_{Kep},
\end{equation}
and the virial factor is,

\begin{equation}
\label{collin2}
f={1\over 4(a^2+\sin^2 i)},
\end{equation}
where $a$ can be interpreted as the  $V_{\rm turbulent}/V_{\rm Kep}$ ratio (Collin et al. 2006), although we prefer to interpret it as the $V_{\rm isotropic}/V_{\rm cylindrical}$ ratio. In Figure \ref{factor} we have represented the virial factor corresponding to different values of $a$ and $i$. According to this Figure, to obtain values of the virial factor $f\sim 0.4$ we need to consider relatively large values of $a$ if $i\lesssim 40$.  From Eq. \ref{collin2} it is possible to derive the probability density function of virial factors, $p(f)$, corresponding to a random inclination of the quasars,

\begin{equation}
\label{PDF}
p(f) = {1\over 8 f^2}{1\over \sqrt{1 - {1\over 4 f} + a^2}},
\end{equation}
{for: $1/(4a^2)\ge f >1/(4(1+a^2))$. From equation \ref{PDF} we can obtain the cumulative probability density function of virial factors,

\begin{equation}
\label{CDF}
c(f) = \int^{f}_{1\over 4(1+a^2)}{1\over 8 f'^2}{df'\over \sqrt{1 - {1\over 4 f'} + a^2}}=\sqrt{1-{1\over4f}+a^2},
\end{equation}
for $f\le1/(4a^2)$. Using a $\chi^2$ criterion, we compare the observed and theoretical cumulative histograms to obtain best-fit estimates of $a$ and $i$. The resulting best-fit models are plotted in Figure \ref{Histograms} for  H$\beta$, H$\alpha$ and MgII. The best-fit values for the inclination, $i$, and the isotropy parameter, $a$, are: $i_{H\beta}\sim{\bf 59}^{\rm o}$, $i_{H\alpha}\sim{ \bf 57}^{\rm o}$,  $i_{MgII}\sim 57^{\rm o}$,  $a_{H\beta}\sim {\bf 0.50}$, $a_{H\alpha}\sim {\bf 0.44}$ and $a_{MgII}\sim {\bf 0.49}$ ({in the case of Mg II we assume $R_{MgII}/R_{H\beta} \sim 2$, see \S \ref{secind}). If we limit the inclination to $i\lesssim 30^{\rm o}$, which is more likely for quasars, we obtain (see Figure \ref{Histograms2}): $i_{H\beta}\sim 30^{\rm o}$, $i_{H\alpha}\sim {\bf 28}^{\rm o}$,  $i_{MgII}\sim {\bf 25}^{\rm o}$,  $a_{H\beta}\sim {\bf 0.68}$, $a_{H\alpha}\sim {\bf 0.62}$ and $a_{MgII}\sim{\bf 0.67}$.} In principle the fits look better when the restriction in the inclination is not applied. However, the improvement in the fits may be only apparent as the spreading of the values of $f$ around the mean value, which is easier to fit when the inclination plays a role, is likely due to experimental errors. In any case, the CDF analysis favors a rather high value for the isotropy parameter, $a\sim {\bf 0.4} - 0.7$.
}

%the $R_{MgII}/R_{CIV} \sim 2$ scaling from Shen et al. (2019) and supposing that $R_{CIV}\sim R_{H\beta}$.}

%We have considered a maximum inclination of $i_0=45^º$, and two values for the parameter a, $a=0.3$ and $a=0.8$. It is clear that even for this large value of the maximum inclination of quasars, we need to consider high values of $a$ (about 0.8) to explain the H$\beta$ and H$\alpha$ virial factors. 

The virial factors associated to the Fe III$\lambda\lambda$2039-2113 blend, $\langle f_{FeIII}\rangle = 14.3\pm 2.4$, are really high as compared with the results for the Balmer lines, although Liu et al. (2017)  find values in the $f\sim8$ to $\sim16$ range using redshifts (interpreted as gravitational) and widths of several emission lines in Mrk 110. The distribution of the Fe III emitters in a flattened structure  with an inclination of a few degrees can explain the large observed values of $f_{FeIII}$, which imply $a\lesssim 0.13$. This is in agreement (Guerras et al. 2013a,b, Fian et al., 2018) with the large microlensing magnifications experimented by the Fe III, comparable to those of the continuum and much larger than those of other emission-lines (like H$\beta$ or CIV), supporting that  the Fe III emission arises from a region related to the accretion disk, while the emitters of H$\beta$ and  H$\alpha$ belong to a larger, less flattened structure.

\section{Conclusions}

Using a new method to measure SMBH masses, based on the redshift of the Fe III$\lambda\lambda$2039-2113 blend, we estimate the virial factor in quasars and study the kinematics of the emission line emitters. The main conclusions are the following:

%{\color{red} The correlation between the FWHM of several emission lines (H$\beta$, H$\alpha$, Mg II, and Fe III) and the redshift of the Fe III$\lambda\lambda$2039-2113 blend, allows us to measure the virial factor and to study the kinematics of the emitters. The main conclusions are the following,}

1 - We have obtained individual measurements of the virial factor in 10 quasars corresponding to the Balmer lines, H$\beta$, H$\alpha$, {and to Mg II}. The mean values for the Balmer lines, $\langle f_{H\beta}\rangle = {\bf 0.43\pm 0.20}$ and $\langle f_{H\alpha}\rangle = {\bf 0.50\pm 0.24}$, agree with previous estimates for lines of comparable widths when the FWHM are derived from the mean spectrum as it is our case. {We obtain similar results for Mg II, $\langle f_{MgII}\rangle \sim{\bf  0.44\pm 0.21}$, scaling the Mg II to C IV according to Shen et al. (2019), $R_{MgII}\sim 2R_{CIV}$,  and CIV to H$\beta$, $R_{CIV}\sim R_{H\beta}$.}

2 - We have also measured the virial factors associated to the Fe III$\lambda\lambda$2039-2113 emitters obtaining very high values, $\langle f_{FeIII}\rangle = 14.3\pm 2.4$, only comparable to other measurements based on gravitationally redshifted lines (Liu et al. 2017). 

{3 - Within the statistical limits imposed by the small size of the sample, there is a relatively narrow range of variation of virial factors for H$\beta$, H$\alpha$, and Mg II on one side, and Fe III on the other. This fact is reflected in the correlations between the Fe III$\lambda\lambda$2039-2113 redshift and the squared emission line widths.  The virial factors directly derived fitting these relationships are very similar to the average of the individual measurements.}  

%{\bf In spite of the scatter of the data, the logarithmic fits are, in all the cases except the  $f_{MgII}^{BOSS}$, consistent with the initial hypothesis: virialization of the BEL and  gravitational origin of the Fe III$\lambda\lambda$2039-2113 redshifts.
%}

4 - {The relatively} small scatter in virial factors not only supports a significant amount of regularity in the geometry and dynamics of the emission line regions among the quasars of the sample (not neccessarily the same geometry and dynamics for different emission line regions), but also that the virial factor might not be the main cause of the scatter found in virial mass determinations.

5 - According to the virial measurements, the BLR region associated to the Balmer lines, {may} be a 3D structure, {with a significant contribution from} isotropic motion. Comparing the observed cumulative histograms of virial factors, $CDF(f)$, with the predictions of a simple model, we find that the ratio between isotropic and cylindrical contributions to the kinematics is rather high, $a\sim {\bf 0.4} - 0.7$. On the contrary, the Fe III$\lambda\lambda$2039-2113 emitters likely belong to an almost face on flattened region, which may be identified with the accretion disk.

With this work we have shown that the method based on the redshift of the Fe III$\lambda\lambda$2039-2113 blend can be successfully used as a primary mass indicator to study not only the SMBH but also the physics of its environment.

\acknowledgements{We thank the anonymous referee for the thorough review of the paper. We thank the SDSS and BOSS surveys for kindly providing the data. This research was supported by the Spanish MINECO with the grants AYA2016-79104-C3-1-P and AYA2016-79104-C3-2-P.  J.J.V. is supported by the project AYA2017-84897-P financed by the Spanish Ministerio de Econom\'\i a y Competividad and by the Fondo Europeo de Desarrollo Regional (FEDER), and by project FQM-108 financed by Junta de Andaluc\'\i a.  V.M. gratefully acknowledges partial support from Centro de Astrof\'\i sica de Valpara\'\i so. C.F. acknowledges support of La Caixa fellowship.}

\appendix
%
%{ \color{blue} 
%\section{New Estimates of the Mg II Widths of the BOSS Quasar Composite Spectra. Correlation between the Fe III$\lambda\lambda$2039-2113 Redshifts and $FWHM_{MgII}^2$. \label{appa} }
%
%In Mediavilla et al. (2019) the $FWHM_{MgII}$ was estimated in a direct way, fitting a straight line to the continuum and measuring the width at 50\% of the peak intensity with respect to this continuum. However, it is commonly supposed that the wings of MgII are strongly contributed by a pseudo-continuum arising from different species, particularly iron. Thus, to refine this calculation,  we fit and remove these wings using two Gaussians, one to match the broad wings and other to account for the comparatively narrow MgII emission\footnote{{\color{blue} More complex approaches to reproduce and remove the Mg II broad wings are out of the scope of this paper.}}. The correlation of the Fe III$\lambda\lambda$2039-2113 redshifts with the FWHM of the narrower Gaussian that we identify with the MgII emission is even better (see Fig. \ref{Correlations}) than in  Mediavilla et al. (2019). The data points (with a 14\% of global scaling) match perfectly with the results (and linear fit) obtained from Mej{\'{\i}}a-Restrepo et al. (2016) data.} 
%

\section{Line Profile for an Isotropic Spherical Shell of Emitters\label{appb}}

We suppose that the emitters are confined to a spherical shell and move tangent to the surface with velocity,

\begin{equation}
\vec v=|\vec v|\cos\alpha\ \vec e_\theta+|\vec v|\sin\alpha\ \vec e_\phi,
\end{equation}
where $\alpha$ is uniformly distributed between $0$ and $2\pi$. The line profile for an observer located at $z=\infty$ can be obtained from,
 
 \begin{equation}
 \label{F}
 F_\lambda\propto \int_0^{2\pi}d\alpha \int_S dS\,\delta\left(\lambda-\lambda_0\left(1+{\vec v\cdot \vec e_z\over c}\right)\right),
 \end{equation}
where $S$ is the surface of the spherical shell. Taking into account that $\vec v\cdot \vec e_z=-|\vec v|\sin \theta$, the integral over $S$ can be written as,

\begin{equation}
\label{surface}
 \int_0^{2\pi} d\phi \int_{-\pi/2}^{+\pi/2} d\theta \sin \theta\, \delta\left(\lambda-\lambda_0\left(1-{|\vec v|\over c}\cos\alpha\sin\theta\right)\right).
 \end{equation}
Defining, $f=\lambda-\lambda_0\left(1-{|\vec v|\over c}\cos\alpha\sin\theta\right)$ and performing the integration over $\phi$, Eq. \ref{surface} can be written,

\begin{equation}
\label{surface2}
{2\pi}\int {df\over \lambda_0{|\vec v|\over c}\cos\alpha\cos\theta}  \sin \theta\, \delta\left(f\right).
 \end{equation}
This integral is null except when $f=0$, that is, $\sin\theta={x\over \cos\alpha}$, which implies, ${x\over \cos\alpha}\le1$, where we have defined, $x={c\over |\vec v|}{\lambda-\lambda_0 \over \lambda_0}$. Thus we can integrate Eq. \ref{surface2} to obtain,

\begin{equation}
\label{result}
{2\pi\over \lambda_0{|\vec v|\over c}}{{x \over \cos\alpha}\over \sqrt{\cos^2\alpha-x^2}},
 \end{equation}
with the condition, ${x\over \cos\alpha}\le1$. Now, Eq. \ref{F} can be written,

\begin{equation}
 F_x\propto \int_0^{\arccos x}d\alpha {2\pi\over \lambda_0{|\vec v|\over c}}{{x \over \cos\alpha}\over \sqrt{\cos^2\alpha-x^2}},
\end{equation}
with $x\le 1$ and $F_{-x}=F_x$. The result of this integral is a constant,

\begin{equation}
 F_x\propto {\pi^2\over \lambda_0{|\vec v|\over c}},
\end{equation}
for $-1\le x\le 1$. Thus, the resulting profile is a rectangular function of width $2|\vec v|$.

\clearpage

\begin{figure}[h]
%\vskip -1 truecm
\includegraphics[scale=0.75]{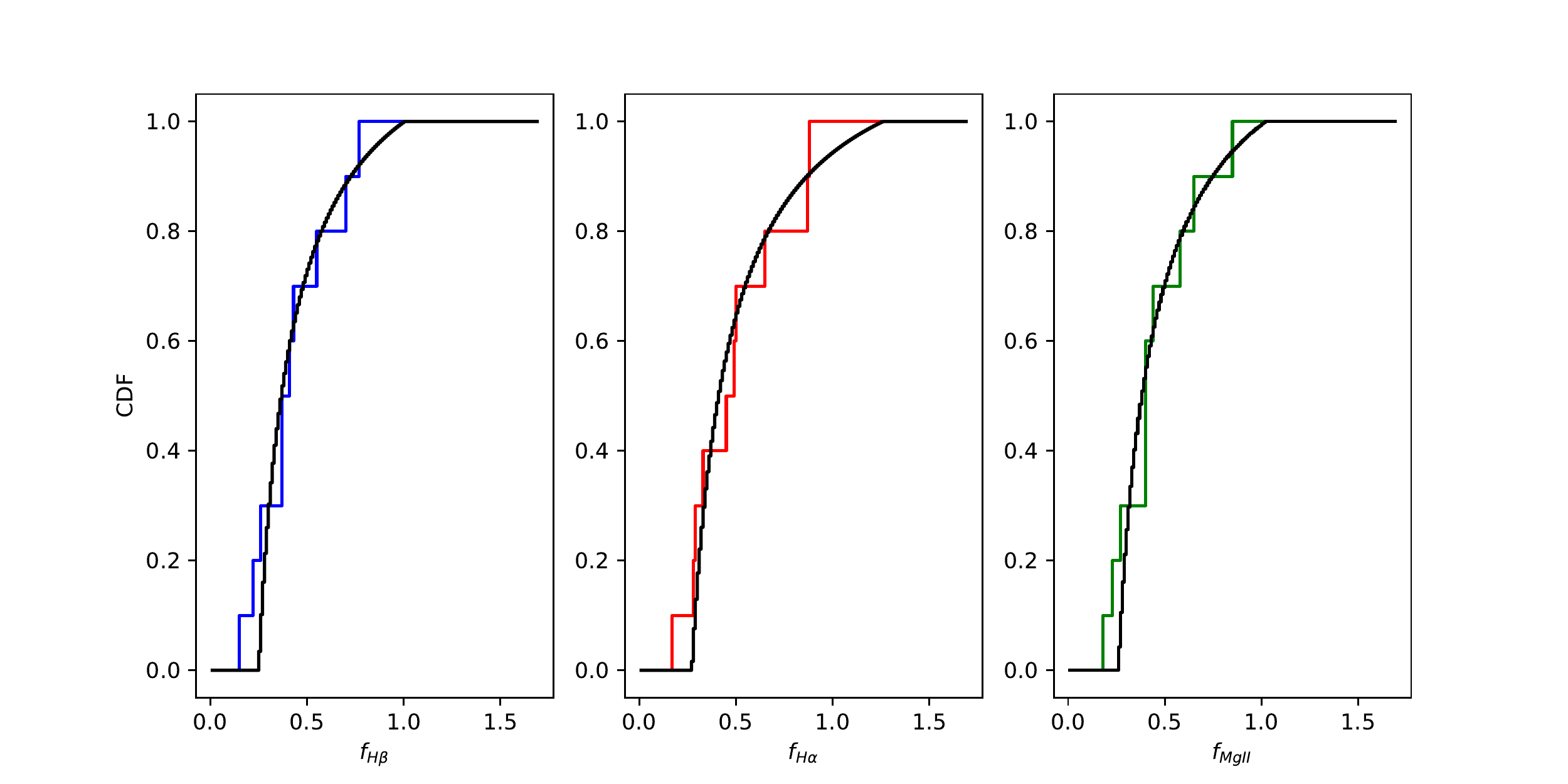}
%\plotone{fit3_SDSS.pdf}
\caption{Cumulative distributions of the virial factors, $CDF(f)$, corresponding to H$\beta$, H$\alpha$ and MgII. The continuous lines correspond to the theoretical best fit models (see text).\label{Histograms}}
\end{figure}

\begin{figure}[h]
%\vskip -1 truecm
\includegraphics[scale=0.75]{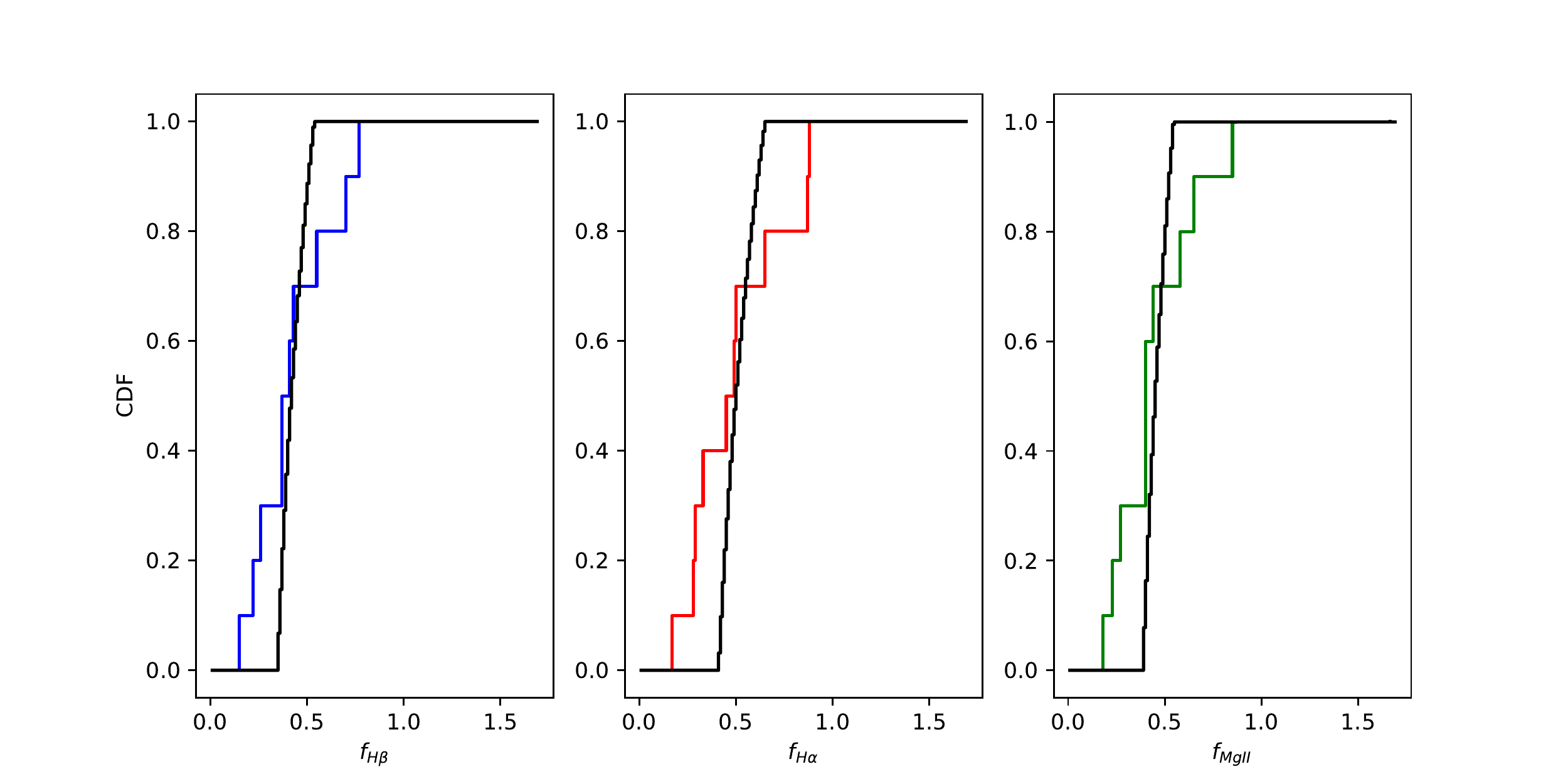}
%\plotone{fit3_SDSS.pdf}
\caption{Cumulative distributions of the virial factors, $CDF(f)$, corresponding to H$\beta$, H$\alpha$ and MgII with the restriction $i\le 30^{\rm o}$. The continuous lines correspond to the theoretical best fit models (see text).\label{Histograms2}}
\end{figure}

\begin{figure}[h]
%\vskip -1 truecm
\includegraphics[scale=0.8]{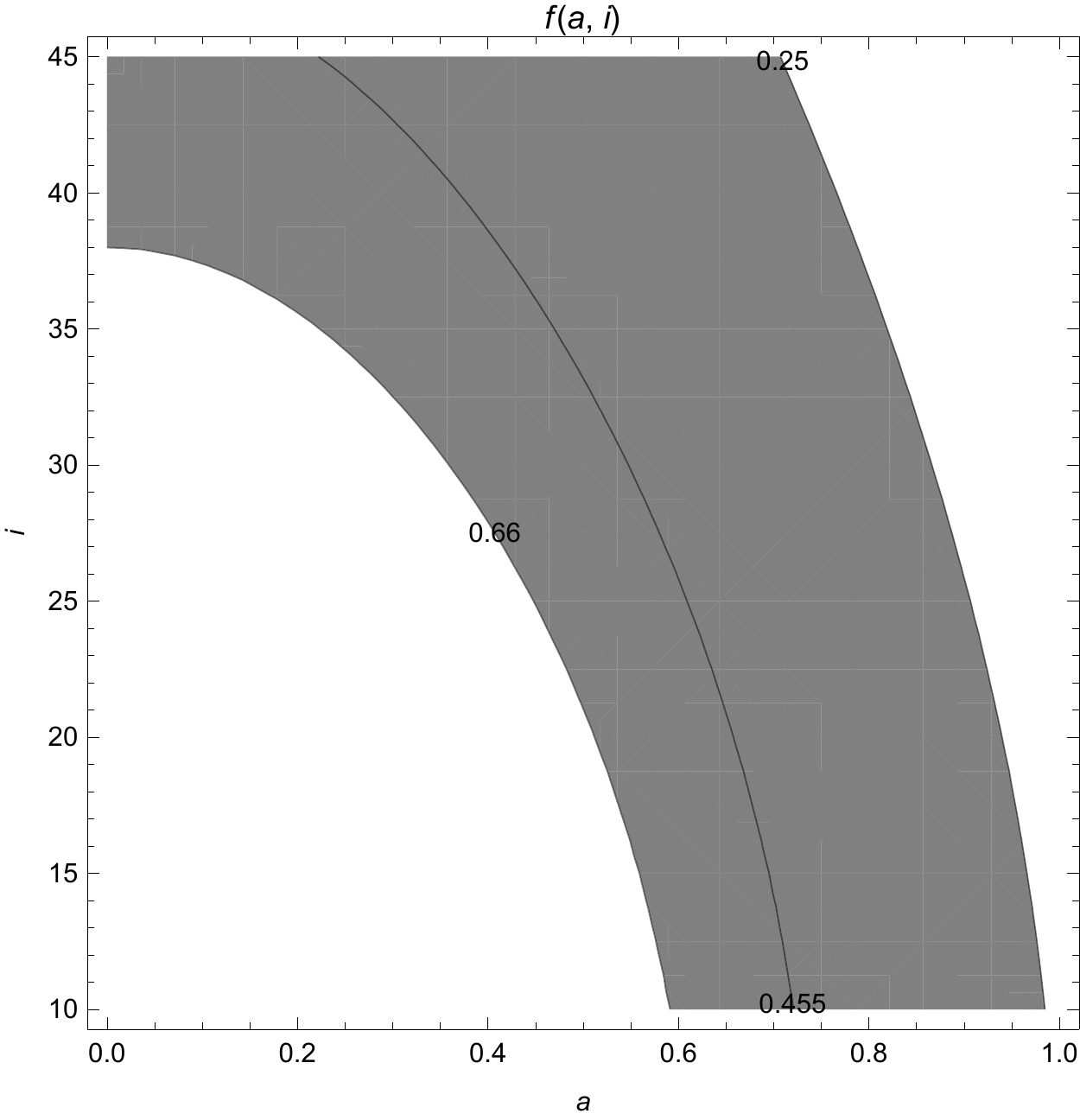}
%\plotone{fit3_SDSS.pdf}
\caption{Virial factor, $f(a,i)$,  as a function of the isotropy parameter, $a$, and the inclination of the quasar, $i$ (see text). The shaded region corresponds to the average of the virial factor for the Balmer lines $\pm1\sigma$.\label{factor}}
\end{figure}

\clearpage
\clearpage

\begin{table}
%\centering
\caption{Virial Factors for 10 Quasars}
\medskip
\begin{tabular}{cccccc}
\hline
Quasar &$\log( \lambda L_{\lambda1350}/\rm erg\, s^{-1})$&$\log( \lambda L_{\lambda5100}/\rm erg\, s^{-1})$&$f_{H\beta}$ & $f_{H\alpha}$ & $f_{MgII}$ \\
\hline
J0019$-$1053& 45.90&45.40&$0.70\pm 0.15$ & $0.66\pm 0.16$ & $0.59\pm 0.10$\\
J0043$+$0114& 46.48&45.93&$0.56\pm 0.18$ & $0.88\pm 0.20$ & $0.40\pm 0.09$ \\
J0155$-$1023& 46.64&46.13&$0.22\pm 0.05$ & $0.29\pm 0.06$ & $0.27\pm 0.06$ \\
J0209$-$0947& 46.58&46.09&$0.42\pm 0.08$ & $0.49\pm 0.09$ & $0.65\pm 0.11$\\
J0404$-$0446& 45.92&45.62&$0.44\pm 0.58$ & $0.46\pm 0.41$ & $0.44\pm 0.40$ \\
J0842$+$0151& 46.41&45.79&$0.26\pm 0.11$ & $0.29\pm 0.11$ & $0.23\pm 0.08$ \\
J0934$+$0005& 46.17&45.68&$0.77\pm 0.33$ & $0.88\pm 0.38$ & $0.41\pm 0.17$ \\
J0941$+$0443& 46.29&45.79&$0.38\pm 0.09$ & $0.34\pm 0.07$ & $0.40\pm 0.10$\\
J1002$+$0331& 46.58&45.99&$0.38\pm 0.11$ & $0.50\pm 0.09$ & $0.86\pm 0.17$ \\
J1158$-$0322& 46.54&46.08&$0.15\pm 0.06$ & $0.18\pm 0.07$ & $0.18\pm 0.07$ \\
\hline
\end{tabular}
\end{table}

\begin{table}
%\centering
\caption{Virial Factors for BOSS Quasar Composite Spectra}
\medskip
\begin{tabular}{ccccc}
\hline
Composite \#&$ \log( \lambda L_{\lambda1350}/\rm erg\, s^{-1})$&$\log( \lambda L_{\lambda5100}/\rm erg\, s^{-1})$&$f_{MgII}$ & $f_{FeIII}$ \\
\hline
2&45.35 &45.05 &$0.32\pm 0.04$ & $13.71\pm 1.97$  \\
4& 45.66&45.09&$0.55\pm 0.06$ & $11.58\pm 1.38$  \\
5& 45.68&45.38&$0.33\pm 0.04$ & $14.71\pm 2.02$  \\
6& 45.64&45.62&$0.21\pm 0.02$ & $11.73\pm 1.64$  \\
7& 46.04&45.51&$0.54\pm 0.06$ & $12.71\pm 1.60$  \\
8& 46.06&45.76&$0.32\pm 0.04$ & $12.20\pm 1.61$  \\
9& 46.02&45.99&$0.21\pm 0.02$ & $12.11\pm 1.58$  \\
13& 45.66&45.09&$0.61\pm 0.06$ & $12.91\pm 1.47$  \\
14& 45.68&45.38&$0.33\pm 0.04$ & $15.06\pm 2.09$  \\
15& 45.65&45.63&$0.20\pm 0.03$ & $16.87\pm 3.28$  \\
16& 46.04&45.51&$0.44\pm 0.04$ & $17.48\pm 2.31$  \\
17& 46.06&45.76&$0.26\pm 0.03$ & $11.28\pm 1.55$  \\
18& 46.02&45.99&$0.16\pm 0.02$ & $13.79\pm 2.25$  \\
23& 45.68&45.38&$-$ & $18.04\pm 2.76$  \\
25& 46.05&45.52&$-$ & $18.39\pm 2.48$  \\
26& 46.06&45.76&$-$ & $16.16\pm 2.42$  \\
\hline
\end{tabular}
\end{table}

\end{document}